\documentclass[11pt,preprint]{aastex}
\input psfig.sty

\newcommand {\ie} {{\it i.e.}}
\newcommand {\cf} {{\it cf.}}
\newcommand {\eg} {{\it e.g.}}
\newcommand {\ea} {{\it et~al.}}
\newcommand {\be} {\begin{equation}}
\newcommand {\ee} {\end{equation}}

\def\refitem{\par\parskip 0pt\noindent\hangindent 20pt}

\slugcomment{}
\shorttitle{Comptonization of IR radiation in quasars}
\shortauthors{B{\l}a{\.z}ejowski \ea}

\begin{document}

\title{Comptonization of Infrared Radiation from Hot Dust by Relativistic 
Jets in Quasars}

\author{M.~B{\l }a\.zejowski and M.~Sikora}
\affil{Nicolaus Copernicus Astronomical Center, Bartycka 18, 00-716
Warsaw, Poland}
\email{blazejow@camk.edu.pl}

\author{R.~Moderski\altaffilmark{1}}
\affil{JILA, University of Colorado, Boulder, CO 80309-0440, USA}

\and

\author{G.~M.~Madejski\altaffilmark{2}}
\affil{Laboratory for High Energy Astrophysics, NASA/GSFC, 
Greenbelt, MD 20771, USA}

\altaffiltext{1}{also: Nicolaus Copernicus Astronomical Center,
Warsaw}
\altaffiltext{2}{also: Dept. of Astronomy, University of Maryland,
College Park, MD 20771, USA}

\author{\bf Accepted for publication in the Astrophysical Journal, Dec.  20, 2000 issue}

\begin{abstract}

We demonstrate the importance of near-infrared radiation from hot dust
for Compton cooling of electrons/positrons in quasar jets.  In our
model, we assume that the non-thermal radiation spectra observed in
OVV quasars are produced by relativistic electrons/positrons
accelerated in thin shells which propagate down the jet with
relativistic speeds.  We show that the Comptonization of the near-IR
flux is likely to dominate the radiative output of OVV quasars in the
energy range from tens of keV up to hundreds of MeV, where it exceeds
that produced by Comptonization of the UV radiation reprocessed and
rescattered in the Broad Emission Line (BEL) region.  The main reason
for this lies in the fact that the jet encounters the ambient IR
radiation over a relatively large distance as compared to the 
distance where the energy density of the broad emission
line light peaks.  In the soft - to mid energy X--ray band, the spectral
component resulting from Comptonization of the near-IR radiation joins
smoothly with the synchrotron-self-Compton component, which may be
responsible for the soft X--ray flux.  At the highest observed
$\gamma$--ray energies, in the GeV range, Comptonization of broad
emission lines dominates over other components.

\end{abstract}


\section{INTRODUCTION}

Recent discovery that blazars are strong and variable $\gamma$--ray
emitters \citep{vmon95} provided independent evidence that the blazar
phenomenon (smooth continuum emission in all observable bands;
large-amplitude, rapid variability; and high linear polarization) is
produced by relativistic sub-parsec jets which are oriented at small
angles to the line of sight (Dondi \& Ghisellini 1995;  for a recent 
review, see, \eg, Ulrich, Maraschi, \& Urry~1997).  Radio loud quasars with
such an orientation of the jet are observed as 
optically-violent-variable (OVV) and highly polarized quasars, while
the line-weak radio galaxies are seen as BL Lac objects.

Non-thermal spectra of blazars can be divided into two components, the
low energy component, commonly interpreted in terms of the synchrotron
radiation mechanism, and the high energy component, presumably
produced by the inverse-Compton (IC) process (see, \eg, review by
Sikora~1997).  In BL Lac objects, the IC energy losses of relativistic
electrons/positrons are likely to be dominated by the synchrotron
self-Compton (SSC) process \citep{maki97,ghis98,copaha99}, whereas in
the OVV quasars, Comptonization of the diffuse ambient radiation field
is probably more important (Sikora, Begelman, \& Rees~1994 [hereafter:
SBR94]; Blandford \& Levinson~1995). The above, however, should be
regarded as a trend rather than the rule, and, \eg, the case of BL
Lacertae shows that Comptonization of external radiation can be
important also in BL Lac objects \citep{mad99}.

The models proposing that $\gamma$--rays in OVV quasars result from
Comptonization of the diffuse ambient radiation field are commonly
known as external radiation Compton (ERC) models.  There are many
variants of the ERC models. In some of them, radiation sources are
approximated by homogeneous ``blobs'' propagating along the jet
(SBR94; Ghisellini \ea~1998; Kusunose, Takahara, \& Li~2000), while in
others --- by an inhomogeneous flow \citep{bl95}. They also differ
regarding the dominant diffuse ambient radiation field.  For small
distances, $< 10^{16-17}$ cm, this can be provided directly by the
accretion disc \citep{dersch93}, while at larger distances, the broad
emission lines (BEL) and near-IR radiation from hot dust are likely to
dominate (SBR94). In the ``blob'' model, the distance along the jet
where most of the $\gamma$--ray flux is produced can be estimated from
variability time scale and from the location of the spectral break
where the luminosity of the Compton component peaks.  The $1 \div 3$
day time scales of $\gamma$--ray outbursts and typical location of
their spectral breaks in the $1 \div 30$ MeV range are consistent with
production of $\gamma$--rays by Comptonization of light from BEL
regions and dust at distances $10^{17} - 10^{18}$ cm (see Table~1 in
SBR94). 

The situation in the X--ray band is more complex.  In part, the SSC
process may be important in the soft/mid energy X--ray bands
\citep{ino96,kubo98}.  Furthermore, X--ray variability usually shows
lower amplitude than variability in the $\gamma$--ray band
\citep{sam97,weh98}. This suggests that X--rays produced co-spatially
with $\gamma$--rays are contaminated by another source of X--rays,
possibly located at larger distances along the jet and therefore
varying on much longer time scales.  Superposition of radiation
components produced at lower and higher distances can also explain
progressively weaker variability in the synchrotron component at lower
frequencies \citep{bro89,ede92,har96,umu97}.  Since both sources
located at smaller and larger distances are related to dissipation
events presumably triggered by collisions between inhomogeneities
propagating down the jet at different velocities, one can expect that
location and strength of these collisions change with time.  As a
result, the relative amplitude of different spectral components may
also be changing from epoch to epoch. The best studied example is the
quasar 3C~279, which is usually very variable in the optical band
(this is an OVV quasar), but during the high energy flare in 1996, the
optical flux varied less than 20\% \citep{weh98}.

As was demonstrated by \citet{sima00}, studies of X--ray radiation in
OVV quasars allow a verification of the presence of low energy
electrons in a jet.  This, in turn, can be used to estimate the pair
content of quasar jets. We conduct such studies in this paper by more
detailed modeling of production of radiation.  This is also the first
work where the contribution of all three Compton components: SSC,
C(BEL) and C(IR) is taken into account simultaneously.

This paper is organized as follows: in \S2 we describe our model of
non-thermal flare production in relativistic jets.  In \S3 we use the
model to reproduce typical spectra observed in $\gamma$--ray detected
OVV quasars and illustrate their dependence on the low energy break in
the electron/positron injection function. We use two classes of
models, one where the radiative output is dominated by Comptonization
of IR radiation from hot dust, and another where $\gamma$--ray flux
results from Comptonization of broad emission lines.  In that section,
we also discuss the model parameters and, in particular, calculate the
pair content of the jet plasma, required to provide the observed
spectra.  We summarize our main conclusions in \S4.
  
\section{THE MODEL}

\subsection{Evolution of the Electron Energy Distribution}

In our model, we assume that relativistic electrons/positrons are
enclosed within a very thin shell which propagates along the jet with
relativistic speed.  With this, we can describe evolution of the
electron energy distribution, $N_{\gamma}$, using the following
continuity equation (see, \eg, Moderski, Sikora, \& Bulik~2000)
\be {\partial N_{\gamma} \over \partial r} = - {\partial \over \partial
\gamma} \left(N_{\gamma} {d\gamma \over dr}\right) + { Q\over c\beta \Gamma} ,
\ee
where the rate of electron/positron energy losses is
\be {d\gamma \over dr} = {1 \over \beta c \Gamma} \left(d\gamma \over
dt'\right)_{rad}- {2\over3}{\gamma \over r} , \label{eq2} \ee
$r$ is the distance from the central engine, $\Gamma$ is the bulk
Lorentz factor, $\beta = \sqrt {\Gamma^2 -1}/\Gamma$, $\gamma$ is the
random Lorentz factor of an electron/positron, $Q$ is the rate of
injection of relativistic electrons/positrons, and $dr = \beta c
\Gamma dt'$.  The second term on the right-hand side of 
Eq.~(\ref{eq2}) represents the adiabatic losses for 
two-dimensional expansion of the jet.  Radiative energy losses 
are dominated by:

\noindent
(1) synchrotron radiation,
\be \left(d\gamma \over dt'\right)_S = - {4\sigma_T \over 3 m_e c } u_B' \, 
\gamma^2 , \ee
(2) Comptonization of synchrotron radiation,
\be \left(d\gamma \over dt'\right)_{SSC} = - {4\sigma_T \over 3 m_e c } u_S'
\, \gamma^2 , \ee
and (3) Comptonization of external radiation,
\be \left(d\gamma \over dt'\right)_{ERC} = - {16 \sigma_T \over 9 m_e c } 
(u_{diff(BEL)} + u_{diff(IR)}) \Gamma^2 \gamma^2  , \ee
where $u_B' = B'^2/8\pi$ is the magnetic field energy density, $u_S'$
is the energy density of the synchrotron radiation field, $u_{diff(BEL)}$ 
$\simeq$ ${\partial L_{BEL}/ \partial \ln r} \over 4 \pi \, r^2\, c$
is the energy density of the broad emission line field 
present at the actual distance of 
a source/shell propagating downstream a jet, 
$u_{diff(IR)} \simeq \xi_{IR} \, 4 \sigma_{SB} \, T^4/c$ is the energy
density of the infrared radiation field, $\xi_{IR}$ is the fraction of 
the central radiation reprocessed into near infrared by hot dust, 
and $T$ is temperature of dust assumed to be located at $r_{IR} > r$ 
(see Eq. 17). Note that the 
expression for $u_{diff(BEL)}$ does not include the contribution from 
emission lines produced at much larger or smaller distances than $r$.
This is because for any smooth radial distribution of 
external diffuse radiation field, as is likely to be the case
of BELR in quasars (Peterson 1993), the lines produced
over larger distances ($> r$) provide much lower energy densities, while the
lines produced at smaller distances are very significantly 
redshifted in the shell frame. 
Here, as elsewhere in the
paper, primed quantities denote measurements made in the source
co-moving frame.  Energy density of the synchrotron radiation is given
by:
\be u_S' = {1 \over 2 \pi a^2 c} \int_{\nu_{abs}'}^{\nu_{S,max}'}
\nu' L_{S,\nu'}' \, {\rm d} \ln \nu' \, \ee
where $a$ is the cross section radius of the jet at a distance where
the outburst is produced, $\nu_{abs}'$ is the frequency at which the
optical thickness due to synchrotron-self-absorption is equal to
unity\footnote{We calculate $\nu_{abs}'$ using an approximate formula
given by SBR94 and derived assuming spherical geometry of the source.
For the shell geometry, $\nu_{abs}'$ is larger and angle dependent, 
but never exceeds the value for spherical case by more than 
$\le (a/\lambda')^{2/7}$.},
$h \nu_{S,max}' = (4/3) \gamma_{max}^2
(B'/B_{cr}) m_e c^2$, $B_{cr} \simeq 4.4 \times 10^{13}$ Gauss, and
\be \nu' L_{S,\nu'}' \simeq 
{1\over 2} (\gamma N_{\gamma}) m_ec^2 {\left \vert{\rm d}\gamma \over
{\rm d} t'\right \vert_S} . \label{eq7} \ee
It should be mentioned here that in the case of thin shell geometry,  
the synchrotron
energy density, given by formula (6), is reached throughout the source on much 
shorter time scale
than $a/c$, and, therefore, unlike the case of spherical sources, the
SSC flares are not expected to lag significantly behind the synchrotron
flares. We also note that our assumption about the {\sl comoving} thinness 
of shells is well justified by the detailed studies of shocks produced by
collisions of inhomogeneities moving with different velocities, provided
their Lorentz factors differ by less than a factor 2 (see, e.g., Komissarov 
and Falle 1997).

\subsection{Modeling of the Radiation Spectra}

Since we consider only the contribution to the overall spectra by the
regions of the jet which are optically thin, the observed spectra can
be computed using the following formula
\be \nu L_{\nu}(t) 
\simeq {1 \over \Omega_j} \int\!\!\!\int_{\Omega_j} 
\nu' L_{\nu'}'[r; \theta']
{\cal D}^4  \, {\rm d} \cos \theta \, {\rm d} \phi \, , \label{eq8}
\ee
where 
\be r= {c \beta t \over 1 -\beta \cos{\theta}}  \, , \ee
$ {\cal D} = [\Gamma(1-\beta\cos \theta)]^{-1} \equiv
\Gamma(1+\beta \cos \theta')$ is the Doppler factor,
$\theta$ is the angle between velocity of the shell element and
direction to the observer, $\nu = {\cal D} \nu'$, and $t$ is the
observed time (see Appendix A). The intrinsic synchrotron luminosity,
$\nu' L_{S,\nu'}'$, is given by Eq.~(\ref{eq7}). The intrinsic 
SSC luminosity is (see, \eg, Chiang \& Dermer 1999)
\be \nu'L_{SSC, \nu'}' = {\sqrt 3 \sigma_T \over 8 \Omega_j r^2} \, 
{\nu'}^{3/2}
\int_{\nu_1'}^{\nu_2'} N_{\gamma} \left [\gamma= \sqrt {3 \nu' \over 4
\nu_S'} \right ] L_{S,\nu'}' \, \nu_S'^{-3/2} \, {\rm d} \nu_S' \,
, \ee
where
\be \nu_1' = {\rm Max} \left [ \nu_{abs}'; {3 \nu' \over 4 \gamma_{max}^2}
\right ] \, , \ee
and
\be \nu_2' = {\rm Min} \left [ \nu_{S, max}'; {3 \nu' \over 4
\gamma_{min}^2} \right ] \, .
\ee

In contrast to the synchrotron and SSC emission which is isotropic in
the co-moving frame, the radiation produced by Comptonization of
external radiation is anisotropic, and thus
\be \nu' L_{C(i),\nu'}' [\theta'] \equiv 
4 \pi {\partial  (\nu' L_{C(i),\nu'}') \over \partial \Omega_{\vec n'_{obs}}'}
\simeq {1\over 2} \gamma \, N_{\gamma} \, 
m_e c^2 
\, \left \vert {\rm d} \gamma \over {\rm d} t' \right \vert_{C(i)} [\theta']
\, , \label{eq13} \ee
where
\be \left\vert {\rm d} \gamma \over {\rm d} t'\right \vert_{C(i)} [\theta'] 
\simeq 
{4 \sigma_T \over 3 m_e c} \gamma^2 {\cal D}^2 u_i
 \, , \label{eq14} \ee
and $i$: BEL or IR. The right-hand side of Eq.~(\ref{eq13}) is
calculated for
\be \gamma= \sqrt {\nu' \over {\cal D} \nu_i} =
{1 \over {\cal D}} \sqrt{\nu \over \nu_i} \, . \ee
Anisotropy of the ERC process (Eq.~\ref{eq14}) and its
astrophysical implications have been extensively discussed by
\citet{der95}, and some general comments on the significance of this
anisotropy can be found in \citet{sik97}.

\subsection{Application to Homogeneous ERC Models}

Using the formalism given above, we consider homogeneous ERC models.
Specifically, our goal is to compare two scenarios: in the first case,
$\gamma$--rays are produced by Comptonization of near-IR radiation
(which we subsequently denote as models {\bf A}), while in the second
case, it is due to Comptonization of broad emission line photons
(hereafter models {\bf B}).  Our model assumptions are:

\refitem 
$\bullet$ electron/positron injection function is a power-law, $Q = K
\gamma^{-p}$, for $\gamma_b < \gamma < \gamma_{max}$, and has a low
energy tail, $Q \propto \gamma^{-1}$, for $\gamma <\gamma_b$;

\refitem 
$\bullet$ $\gamma_b$ is lower than the energy of the break,
$\gamma_c$, produced due to cooling effect (SBR94\footnote{Note that
in SBR94 the cooling energy break is denoted by $\gamma_b$, while in
our paper, it denotes the low energy break in the injection
function.}) and, therefore, it is the latter which determines the
position of luminosity peaks of the spectral components C(BEL) and
C(IR);

\refitem 
$\bullet$ electrons/positrons are injected at a constant rate within a
distance range $(r_0; 2r_0)$ and uniformly fill the shell whose radial
width is $\lambda' < r_0/ \Gamma $;

\refitem 
$\bullet$ the shell propagates down the conical jet with a constant
Lorentz factor $\Gamma$. The half-opening angle of the jet is
$\theta_j = 1/\Gamma$;

\refitem 
$\bullet$ the intensity of the magnetic field is $B(r) = (r_0/r) B_0$, the 
energy density of broad emission lines is 
$u_{diff(BEL)} = {(r_0/r)^2} u_{BEL,0}$,
and the energy density of infrared radiation is $u_{IR}=const$;

\refitem 
$\bullet$ the observer is located at an angle $\theta_{obs}
=1/\Gamma$.

With the above assumptions, the model approximates a situation where
the shell containing relativistic plasma is formed due to collision of
two perturbances moving down the jet with different velocities (see,
\eg, SBR94 and Rees \& M\'esz\'aros~1994). The low energy break in the
electron/positron injection function, $\gamma_b$, corresponds to the
characteristic energy of pre-heated electrons/positrons (see, \eg,
Kirk, Rieger, \& Mastichiadis 1998), and the low energy tail below this
break is introduced to mimic a limited efficiency of the pre-heating
process, nature of which is very uncertain \citep{hos92,lev96,clem97}.

Input parameters of the models were chosen to reproduce typical
features of OVV quasar flares: location of the $\gamma$--ray
luminosity peak in the 1 -- 30 MeV range; apparent $\gamma$--ray luminosity in
the range $10^{47-48}$ erg s$^{-1}$; apparent synchrotron luminosity in the
range $10^{46-47}$ erg s$^{-1}$; deficiency of radiation in the
soft/mid X--ray bands; maximum synchrotron frequency, $\nu_{S,max}
\sim 10^{15}$ Hz; and time scale of $\gamma$--ray flares on the order
of days. For broadband spectra of blazars see, \eg, \citet{vmon95} and
\citet{fos98}; for time scales of rapid (daily) $\gamma$--ray
variability see, \eg, \citet{mich94}, \citet{mat97} and \citet{weh98}.

\section{RESULTS AND DISCUSSION}

As it was shown by SBR94, the high energy $\gamma$--rays detected by
EGRET on board {\it CGRO} from OVV quasars can be produced by
Comptonization of broad emission line light as well as by Comptonization of
infrared radiation (see Table~1 in SBR94).  Whereas the broad emission
line flux gained a lot of attention as a good candidate for the
dominant source of seed photons for the inverse Compton process in a
jet, the infrared radiation from dust was largely ignored in the
literature (see, however, Wagner \ea~1995a and references therein).
To the best of our knowledge, this paper is the first where production
of {\sl both} spectral components, C(IR) and C(BEL), is treated
simultaneously.  We discuss the energy density of those components below.  

Regarding the diffuse radiation field on sub-parsec scales 
provided by broad emission line clouds, the line luminosities in radio
loud quasars are typically in the range $10^{44}-10^{46}$ erg s$^{-1}$
\citep{cpg97,caji99}.  Assuming that the scale of the broad emission line
region is on the order of the distance at which non-thermal flares are
produced in a jet, one can find that the corresponding energy density at a 
distance $r = 10^{18} \, r_{18}$ cm is
\be u_{diff(BEL)} \simeq {L_{BEL} \over 4 \pi r^2 c} \simeq 
3 \times 10^{-3} \, {L_{BEL,45} \over r_{18}^2} \, {\rm erg \, s}^{-1} \, , \ee
where $L_{BEL,45} = L_{BEL}/10^{45}{\rm erg \, s}^{-1}$.
For the diffuse IR radiation, 
just as in the cases of Seyfert galaxies and radio quiet quasars, the
spectra of lobe-dominated radio loud quasars show very prominent
near-IR bumps \citep{san89,dev98}.  These bumps are commonly
interpreted as thermal radiation produced by hot dust.  Such dust 
is expected to be heated by UV radiation of an accretion disc
and located  in the innermost parts of a geometrically thick molecular torus, 
at a distance 
\be r_{IR} \simeq \sqrt {L_{UV} \over 4 \pi \sigma_{SB} T^4 } \simeq 
4 \times 10^{18} \, {L_{UV,46}^{1/2} \over T_3^{2}} \,{\rm cm} , \ee
where $L_{UV} = 10^{46} \,L_{UV,46}$ erg s$^{-1}$ is the luminosity of
the accretion disc, and $T = 10^3 \, T_3 $ K is the effective
temperature of dust.  The dust provides radiation field with energy
density which at $r < r_{IR}$ is approximately constant and equal to
\be u_{IR} \simeq \xi_{IR} \, 4 \sigma_{SB} \, T^4/c \simeq 
2.3 \times 10^{-3} \, (\xi_{IR}/0.3) \, T_3^4 \, {\rm erg \, s}^{-1} \, . \ee

The model parameters are given in Table~2, and illustrated in
Figures~\ref{fig1} -- \ref{fig3}, which show time-averaged broad-band
spectra computed using the model equations and assumptions given in
\S2.  Specifically, we consider three cases: (1) a pure C(IR) model
({\bf A}, shown in Fig.~\ref{fig1}); (2) pure C(BEL) model ({\bf B},
shown in Fig.~\ref{fig2}); and (3) a combination of the two ({\bf
A+B}, shown in Fig.~\ref{fig3}).  In all three figures, a comparison
of the four panels illustrates the dependence of the time averaged
spectra on the low energy break in the injection function, $\gamma_b$.

To validate the models in greater detail, we also consider spectra
obtained for the two well-studied objects.  In Fig.~\ref{fig4} we
present our model fits to the Jan-Feb 1996 outburst in 3C~279
\citep{weh98} and to the March 1995 outburst in PKS~0528+134
\citep{bla97,kubo98}, using a combination of models {\bf A} and {\bf
B}. The model parameters for those fits are presented in Table~2.  

\subsection{Spectra}

Figures~\ref{fig1} -- \ref{fig3} clearly illustrate that the general
features of spectra of $\gamma$--ray detected OVV quasars -- in
particular, the high ratio of Compton to synchrotron luminosities, and
the comparative deficit of X--ray radiation -- can be reproduced by
both types of models, {\bf A} (where C(IR) dominates) and {\bf B}
(where C(BEL) dominates).  However, the details of the spectra are
different.  The models {\bf A} yield X--ray spectra which are much
smoother and closer to the observed ones than the spectra produced by
models {\bf B}. This results from the fact that in order to reproduce
the observed spectra, in models {\bf A}, the Lorentz factors of
electrons/positrons $\gamma_c$ (where the cooling break occurs) need
to be larger than in models {\bf B}, causing the X--ray portion of the
SSC component to be harder in models {\bf A} as compared to {\bf B}.

Another important difference between these two models is in the high
energy $\gamma$--ray spectral band. The spectral components due to
C(IR) have a break at $\sim 1$ GeV, whereas EGRET/{\it CGRO}
observations show that blazar spectra extend at least up to 5
GeV. Therefore, the EGRET spectra can be reproduced only as a
superposition of C(IR) and C(BEL) (see Fig.~3).  This also provides
an attractive explanation for $\gamma$--ray spectra being steeper than
the X--ray spectra by more than the value predicted by cooling effect,
\ie, $\Delta \alpha = 0.5$.  Examples of spectra of objects with
$\Delta \alpha > 0.5$, often called ``MeV blazars,'' can be found in
\citet{mcnb95}, \citet{blom95}, and \citet{mal99}.  

\subsection{Low Energy Injection Break and Pair Content}  

Total number of relativistic electrons plus positrons injected in the
shell during the flare is
\be N_e = \Delta t' \int_{1}^{\gamma_{max}} Q\, d\gamma \,
\simeq {r_0 \over c \Gamma} {K \over \gamma_b^{p-1}} 
\left(\ln \gamma_b + {1\over p-1} \right ) \, , \label{eq16} \ee
where $\Delta t' = \Delta r /c \Gamma = r_0/c \Gamma$ is the injection
time interval. The number of protons enclosed in the shell is on the
order of 
\be N_p \simeq \dot N_p \Delta t_{\lambda} \simeq
{L_p \over m_p c^2 \Gamma} {\lambda \over c} \, ,  \label{eq17} \ee
where $L_p$ is the energy flux of cold protons in a jet and $\lambda$
is the width of the shell as measured in the external frame.  We note here 
that because the energy dissipated in a jet in our model 
results from collisions between inhomogenities, one should not
expect heating of protons to much larger average energies than mildly
relativistic. Furthermore, this energy is quickly lost due to
adiabatic expansion and/or converted back to the bulk energy. This is why 
jet energy flux is expected to be dominated by cold protons.  

Combining Eqs.~(\ref{eq16}) and (\ref{eq17}), we find
\be {N_+ \over N_p} \simeq {N_e \over 2 N_p} \simeq
{m_p c^2 \over 2 L_p} {r_0 \over \lambda} 
{K \over \gamma_b^{p-1}} 
\left(\ln \gamma_b + {1\over p-1} \right ) \, . \ee
(We defined $N_e$ as a sum of numbers of both relativistic electrons and 
positrons, and thus for $N_+ > $ a few $\times N_p$
the charge conservation gives $N_e \sim 2N_+$.)  
Assuming the kinetic luminosity of the jet to be 
$L_j = 10^{47} \, L_{j, 47}$ $\sim 10^{47}$ erg s$^{-1}$ (see, e.g., 
Rawlings \& Saunders 1991; Celotti \& Fabian 1993) 
and noting that 
$\lambda < r_0/\Gamma^2$ (see assumption about the shell width in \S2),
we obtain results presented in Table~\ref{tab4}.

The pair content implied by our fits to the time averaged spectrum of
the outburst in 3C~279 is $n_+/n_p \simeq 4$, and for the outburst in
PKS~0528+134 --- $n_+/n_p \simeq 17$. Note, however, that our fits are
not unique, since such parameters as time scales of flares, energy
densities of external radiation fields, and jet powers 
are not very well constrained by observations.

As we can see from Table~\ref{tab4}, the pair content ranges from a
few for $\gamma_b \sim 100$ up to more than hundred for $\gamma_b
\sim 1$.  The recent detection of circular polarization in several
compact radio sources \citep{war98,howa99} and its interpretation in
terms of the Faraday conversion process favor low values of
$\gamma_b$, and therefore a large pair content. However, there are
some objects with such hard X--ray spectra \citep{mal99,fab98}, that
they can only be explained if $\gamma_b \gg 1$. But even in these
cases, as long as $\gamma_b < 100$, at least some pair content is
required.  Such pairs can be produced via interaction of a cold
proto-jet with radiation of the hot accretion disc corona
\citep{sima00}.  

\subsection{External Radiation Fields}

As we can see from Table~\ref{tab1}, in the models {\bf A} and {\bf
A+B}, $u_{IR} \sim 10^{-4}$erg s$^{-1}$, \ie\ about 20 times less than
energy density of radiation produced by silicate grains with its
maximum temperature $\simeq 1000$ K, and about 100 times less than
energy density of radiation produced by graphite grains with its
maximum temperature $\simeq 1500$ K.  This implies the value of
$u_{IR}$ to be lower than expected, in order to avoid overproduction
of X--rays.  In either model, the corresponding dust temperature is
$\sim 500$ K, \ie\ 2-3 times lower than maximum, and the distance of
dust from the central source is $\sim 5 \sqrt{ L_{UV,46}}$ pc.
Alternatively, overproduction of X--rays by Comptonization of IR
radiation can be avoided by postulating that the $\gamma$--ray
luminosity peak is not defined by the value of the cooling break
$\gamma_c$ as assumed above, but rather, by the injection break
$\gamma_b$ (see Ghisellini \ea~1999; Mukherjee \ea~1999).  

In the model {\bf A+B}, $u_{diff(BEL)} \simeq u_{diff(IR)} \simeq 10^{-4}$ 
erg cm$^{-3}$ 
(see Table~\ref{tab1}), which is 30 times lower than the average in radio
loud quasars.  However, as recent reverberation campaigns show, broad
emission lines have peak luminosities at distances $ \sim 3 \times
10^{17} \sqrt {L_{UV,46}}$ cm (see review by Peterson 1993), and
therefore the amount of the diffuse line radiation at distances where
flares are produced ($\sim r_0 \div 2 r_0 \sim 10^{18}$ cm) is likely
to be a small fraction of the total $L_{BEL}$.  In case of models {\bf
B}, $u_{diff(BEL)}$ is much closer to the observed values and there is no
need for significant reduction of $L_{BEL}$ at distances corresponding
to $r_0$.

\subsection{Bulk Compton Radiation}

The models presented above don't take into account the so-called
bulk-Compton process -- the Comptonization of ambient diffuse
radiation by cold electrons/positrons in a jet \citep{besi87,sik97ea}.
This process should lead to the production of narrow bumps at energies
$\Gamma^2 h\nu_{ext}$, \ie, $\sim 2 \times (\Gamma/15)^2$ keV and
$\sim 0.3 \times (\Gamma/15)^2$ keV for $\nu_{ext} = \nu_{UV}$ and
$\nu_{ext} = \nu_{IR}$, respectively.  The largest contribution should
come from very small distances, where the jet just starts to be
relativistic and collimated.  The absence of soft X--ray bumps in the
observed spectra of OVV quasars (see, \eg, Lawson \& M$^c$Hardy~1998)
suggests that the region of jet formation is very extended, possibly
reaching distances not very much smaller than those where most of
blazar radiation is produced, or else, the number of pairs in a jet is
very small (\cf\ Sikora \ea~1997).  Noting that the number flux of
cold pairs at $r < r_0$ should at least be equal to the number flux of
relativistic pairs accelerated at $r \sim r_0$, the latter condition
corresponds to $\gamma_b \gg 1$.

\subsection{Variability}

Fig.~\ref{fig5} shows the time evolution of the broad-band spectrum
presented in Fig.~\ref{fig3}, binned in the time intervals $\Delta t
= t_0 \equiv r_0 / 2 \Gamma^2 c$.  As one can see from those Figures,
the flares decay at different rates in different spectral
bands. Despite the transverse size of the source and related light
travel effects, the dependence of energy losses on electron/positron
energy is manifested very strongly, causing the synchrotron and
Compton spectra above the luminosity peak to steepen with time.
Additional steepening of the $\gamma$--ray spectrum is caused by the
fact that C(BEL) drops faster with distance than C(IR).  This is
because the energy density of broad emission line light decreases with
distance, while energy density of IR radiation is roughly constant.
Such a steepening of $\gamma$--ray spectra is observationally
confirmed both by direct observations of the slope changes during
individual flares \citep{muc96,muk96,mat97}, and statistically, by
comparison of time averaged spectra as measured during different
epochs \citep{pohl97}.

Steepening of the synchrotron spectrum is also consistent with
observations, which show that the amplitude of variability increases
with the increasing energy of the observing band.  However, in many
OVV quasars, the amplitude of the variability of the synchrotron
component, even at the highest frequencies, is much smaller than
amplitude of variability in $\gamma$--rays.  A particularly clear
example of such behavior is provided by the simultaneous multi-band
observations of the Jan-Feb 1996 flare in 3C~279 \citep{weh98}.  This
can be explained by assuming that the synchrotron component is heavily
contaminated by radiation produced at larger distances and/or by
stationary shocks.

The situation at the keV energies is quite complex, but very
interesting, as it provides important constraints on models.  If the
spectra measured in this range are entirely dominated by ERC
components, then the drop of flux is predicted to be achromatic (with
constant slope). This is because the low energy tails of ERC
components are produced by electrons/positrons whose energy losses are
dominated by adiabatic expansion.  However, as suggested by
\citet{ino96} and \citet{kubo98} and confirmed by our models, the low
energy X--rays are likely to be dominated by the SSC component.  Then
the whole X--ray spectrum is a superposition of a softer SSC component
and harder ERC component, which should result in a concave shape.
Such spectra are in fact in a few cases observed directly, and are
also implied by the fact that two-point slopes between the X--ray and
$\gamma$--ray bands are harder than the X--ray spectra measured in the
soft/mid X--ray bands alone \citep{com97}.  As we can see from
Fig.~\ref{fig5}, such X--ray spectra -- formed by a superposition of
SSC and ERC components -- harden as the source fades.  This is caused
by the SSC component, with luminosity which is proportional to the
square of synchrotron luminosity, decaying faster than the low energy
tail of ERC components.  Hardening of X--ray spectra during the decay
of the flare was in fact observed in some objects (see Ghisellini
\ea~1999; Malizia \ea~1999). In most cases, however, the X--ray
spectrum softens as the flare fades \citep{lhm99,com97}.  This
discrepancy can be resolved by postulating that the X--ray spectra,
particularly at the lowest energies, are contaminated by radiation
produced either at larger distances (see, \eg, Unwin \ea~1997), or in
stationary shocks which may form if jet is reconfined and/or bent (see
Komissarov \& Falle 1997 and references therein).  Furthermore, the
light curves of outbursts can be further modified by a change of shell
velocity (see, \eg, Dermer \& Chiang 1998), or by a change of
direction of propagation. The latter case seems to be favored by VLBI
observations, which often show strong curvature of jets in OVV quasars
on parsec scales \citep{asv96,bow97,ran98,tme98}.  
Finally we emphasize that even very high amplitude flares 
are never isolated:  observations show that they smoothly join with 
lower amplitude neighboring flares, forming with them longer lasting 
high states. In order to explain the continuously flaring high state
light curves in terms of 
propagating shells of relativistic plasma, one must assume that number
of shells per a distance range $r_0 - 2 r_0$ is on the order
of $N_{sh} \sim \Gamma^2 $. For this particular number, 
the number of shells contributing to the observed radiation at a given
instant is $N_{sh,obs} \sim N_{sh}/\Gamma^2 \sim 1$ 
(see Appendix in Sikora et al. 1997). For a much lower
number of shells, the observer would see very isolated
flares, while for much larger number of shells the resultant 
fluctuation (amplitude of the smaller flares superposed on the ``high 
state'') would be smaller than is commonly observed.


\section{CONCLUSIONS}

\refitem $\bullet$ Comptonization of infrared radiation produced by 
hot dust should be taken into account in all radiation models of
blazars.  Such radiation, likely to be present in quasar cores on
parsec scales, is sufficiently dense to compete with broad emission
line flux as the source of Compton cooling of relativistic
electrons/positrons in parsec/sub-parsec jets;

\refitem $\bullet$ Both types of ERC models, either those where the 
$\gamma$--ray luminosity is dominated by C(IR) (models {\bf A}) or
those where the $\gamma$--ray luminosity is dominated by C(BEL)
(models {\bf B}), are able to reproduce all basic broadband spectral
features of OVV quasars during their outbursts;

\refitem $\bullet$ If the blazar radiation is produced 
at distances larger than
the distance at which the energy density of the broad emission line
light peaks, the C(IR) component is expected to dominate over the
C(BEL) component;  

\refitem $\bullet$ The ERC models predict the general steepening of the 
$\gamma$--ray and synchrotron spectra with energy, in a qualitative
agreement with observations.  However, the lower observed variability
amplitude of the synchrotron flux as compared with the $\gamma$--ray
flux suggests a strong contamination of synchrotron radiation produced
in the process of a flare by synchrotron radiation produced at larger
distances or in stationary shocks;

\refitem $\bullet$ X--ray spectra in OVV quasars consist of SSC and ERC 
components, and the latter can be either C(IR) or C(BEL).  Just as in
the case of synchrotron radiation, additional contribution to the
soft/mid energy X--rays is expected from quasi-steady component
produced by distant and/or stationary shocks.  This can explain why
the amplitude of variability in the X--ray bands is smaller than in
$\gamma$--rays;  

\refitem $\bullet$ The ERC models of $\gamma$--ray production predict the 
pair content of the jet plasma, $n_{+}/n_p$, to be in the range from a
few $\times$ $L_{j,47}$ for $\gamma_b \sim 100$ up to hundred $\times$ 
$L_{j,47}$ for $\gamma_b \sim 1$.  Low 
values of $\gamma_b$ (and therefore a large e$^+$/e$^-$ pair content)
are favored by observations of circular polarization, assuming its
interpretation in terms of the Faraday conversion process is correct.

\acknowledgments

The project was partially supported by Polish KBN grant 2P03D 00415,
ITP/NSF grant PHY94-07194, NASA grants NAG-6337 and NASA observing/ADP
grants to the University of Maryland and USRA.

\appendix

\section{RELATIVISTIC TRANSFORMATION OF OBSERVED LUMINOSITY}

\noindent {\sl (1) The case of a point source of radiation}

\bigskip

Observed monochromatic luminosity of a point source at the moment $t$
is
\be \delta L_{\nu} [t] \equiv 
4 \pi {d P_{\nu} \over d \Omega } [t]
=   4 \pi {\cal D}^3 {d P_{\nu'}' \over d \Omega'} [t']
\equiv {\cal D}^3 \, \delta L_{\nu'}' [r; \vec n_{obs}'] 
 \, , \ee
\citep{ryli79}, where 
\be r =  {c \beta t \over 1 - \beta \cos{\theta} } \, , \label{a2}
\ee
\be {\cal D} = {1 \over \Gamma(1-\beta \cos{\theta})} 
\equiv \Gamma (1 + \beta \cos{\theta'}) \, , \ee
$\nu' = \nu/ {\cal D}$, and $\theta$ is the angle between the
direction of the source motion and the direction to the observer,
$\vec n_{obs}$.
 
\bigskip
\noindent {\sl (2) The case of an extended source of radiation}
\bigskip

Since we are only interested in the optically thin radiation, the
observed luminosity of an extended source can be computed by summing
contribution from its small pieces, treated as point sources. Of
course, because of light travel effects, for pieces moving along
different $\theta$'s, contribution must be taken from different
distances (see Eq.~\ref{a2}).  Using approximation of infinitely
thin shell ($\lambda \ll r_0/\Gamma^2$), we divide the source into
$\delta \Omega_j = \Omega_j/ N$ pieces, where N must be enough large
to allow treat individual pieces as point sources (\ie, with
negligible light travel effects).  The observed luminosity is then
given by
\be (\nu L_{\nu})[t] = \Sigma_N \, (\nu \delta L_{\nu})[\theta; t] =
\Sigma_N {\cal D}^4 
(\nu' \delta L_{\nu'}') [r; \vec n_{obs}'] 
 \, . \label{a4} \ee
For uniform shells, one can introduce the formal quantity, ``intrinsic
luminosity'' $L_{\nu}'[\theta; t] = \delta L_{\nu}'[\theta; t] \times
N = \delta L_{\nu}' [\theta; t] \times (\Omega_j /\delta \Omega_j)$,
and with this and $N \to \infty$, Eq.~(\ref{a4}) can be converted
into Eq.~(\ref{eq8}).

\clearpage

\centerline {\bf FIGURE CAPTIONS}

\figcaption[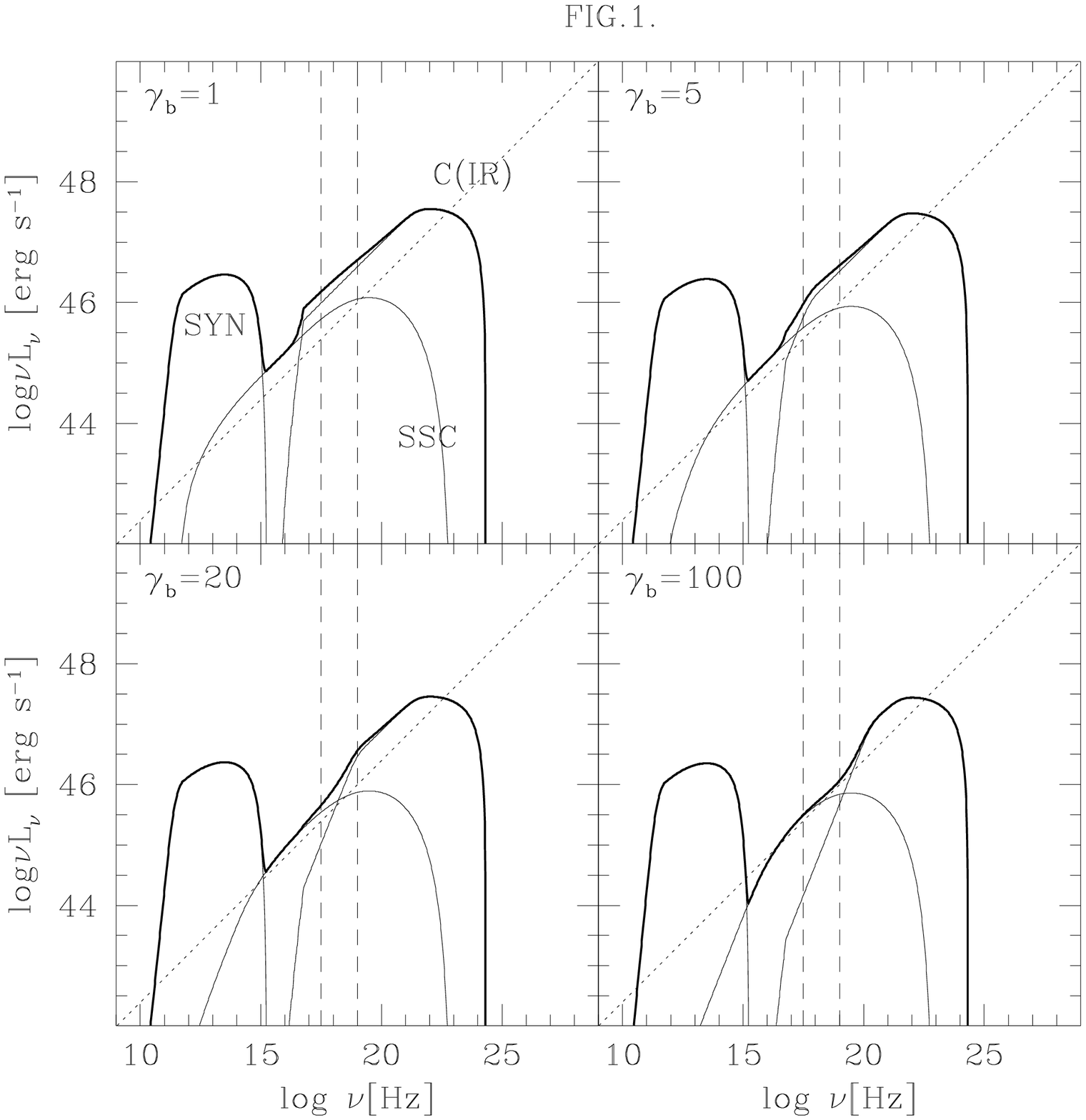]{Time averaged model spectra of blazars for four values of
$\gamma_b$. The dotted line marks the typical slope of the X--ray
spectra observed in OVV-type blazars ($\alpha=0.6$); the dashed lines
enclose the 1--30 keV X--ray band.  In each panel we show three
spectral components: synchrotron (SYN), synchrotron-self-Compton (SSC)
and Comptonization of near-IR dust radiation [C(IR)]. (Model A).  (For
the model parameters, see Table~\ref{tab1}). \label{fig1}}

\figcaption[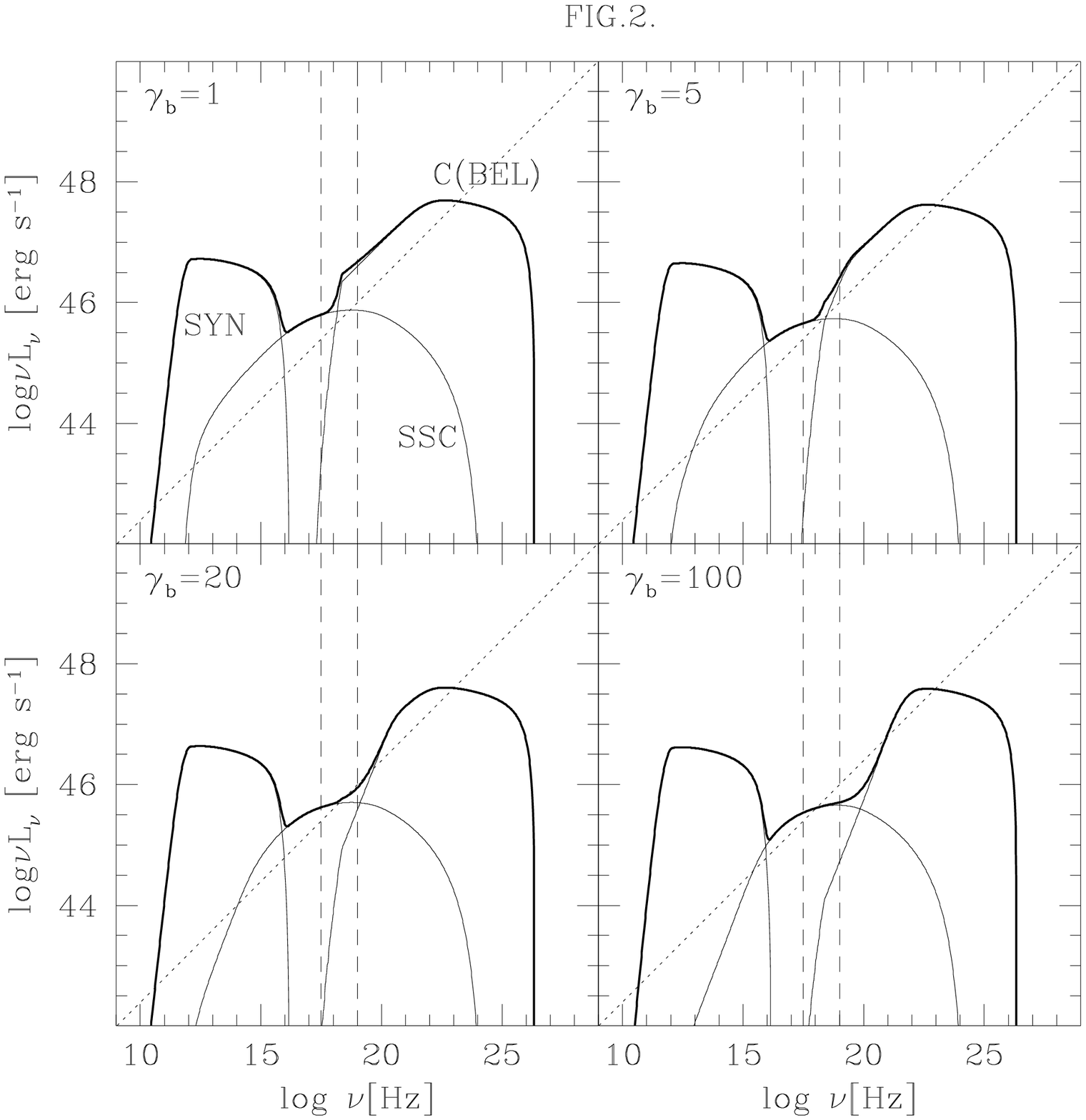]{Same as Fig.~\ref{fig1}, but C(IR) is
replaced with C(BEL) (Model B). \label{fig2}}

\figcaption[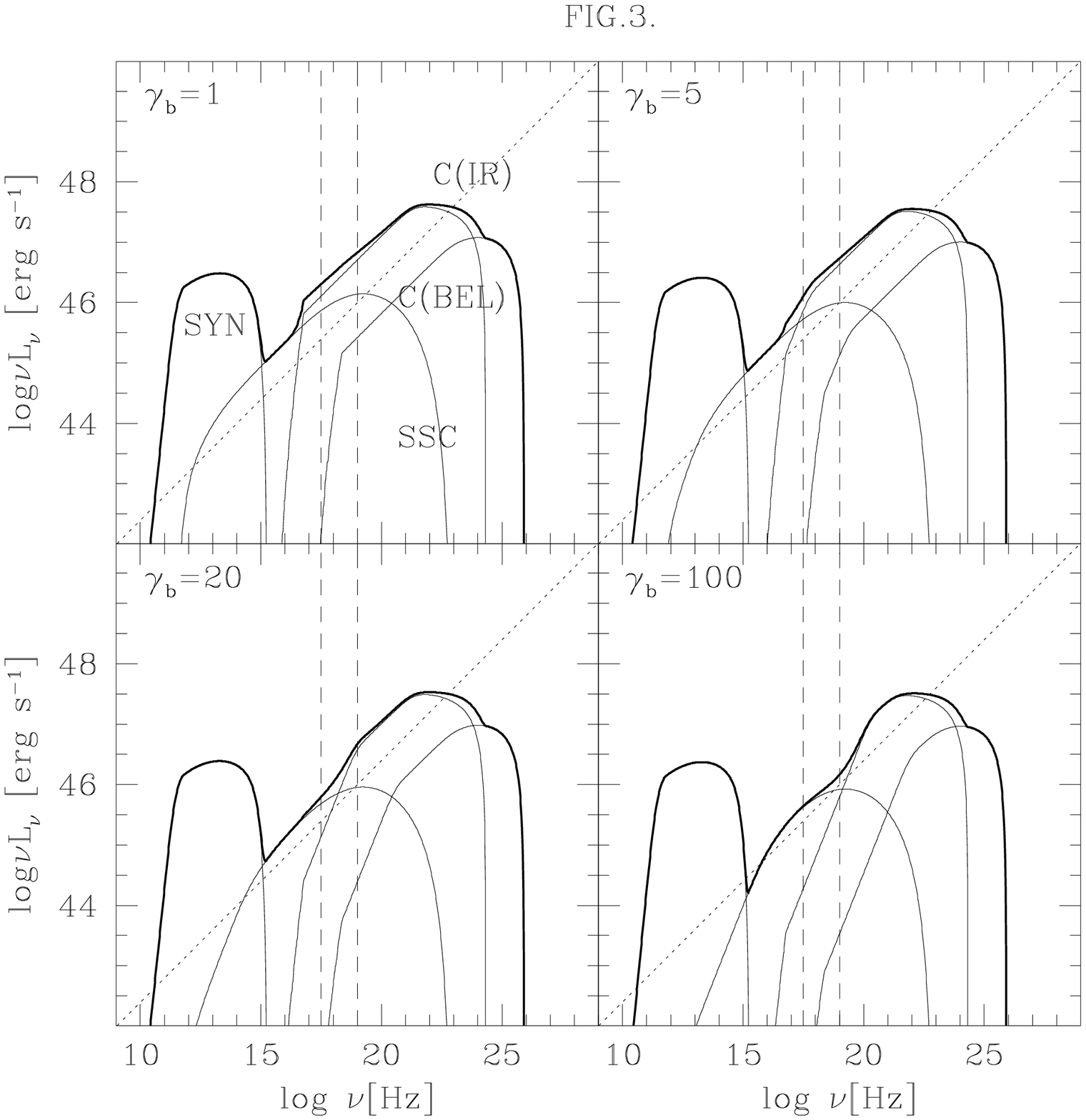]{Same as Fig.~\ref{fig1}, where both C(IR) and C(BEL)
are included (Model A+B). \label{fig3}}

\figcaption[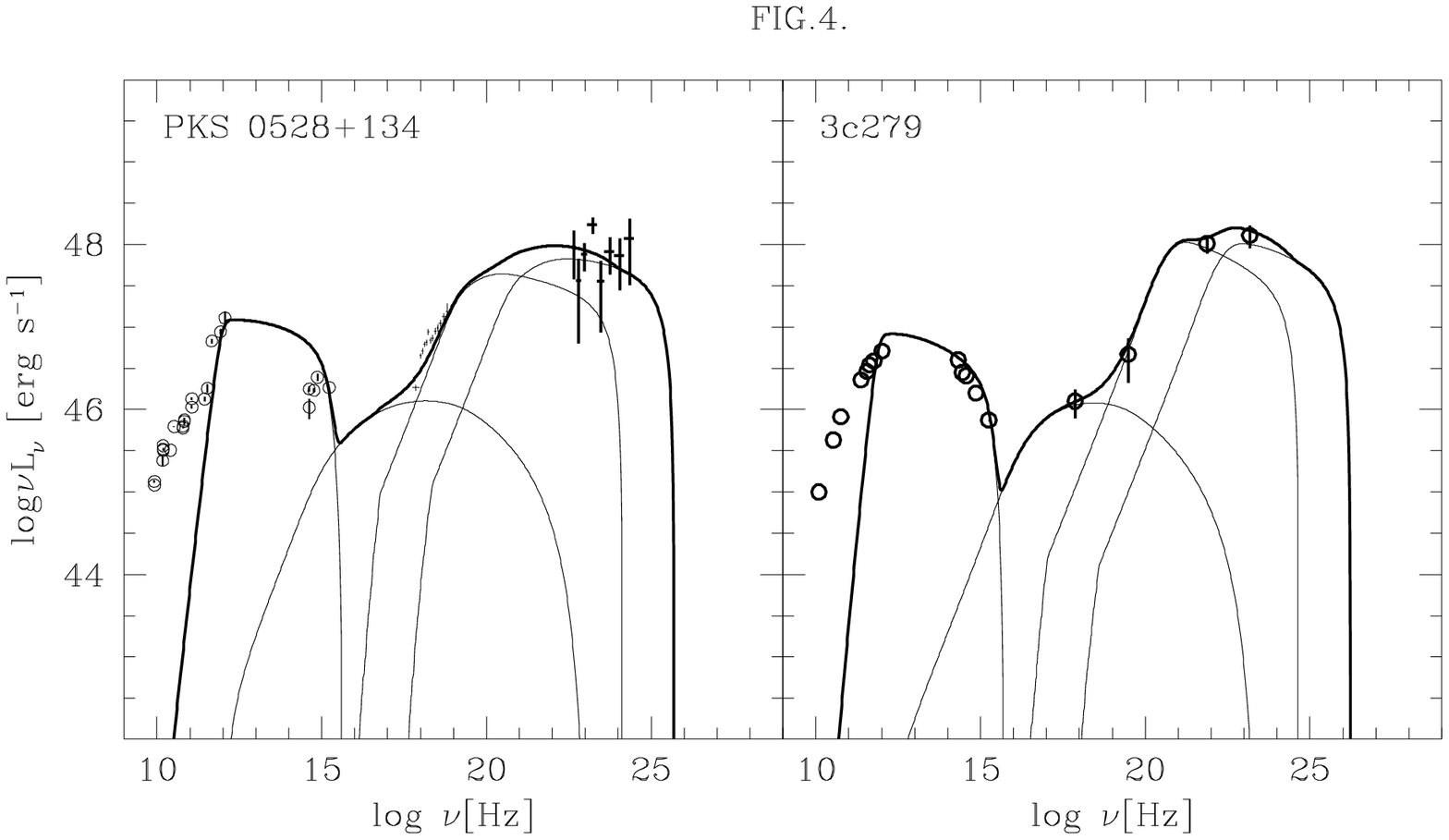]{Left panel: model fit to the outburst spectrum 
observed in 
PKS~0528+134 (Model A+B).  The X--ray spectra (ASCA) and
$\gamma$--rays (EGRET) are simultaneous \citep{mcnb95,kubo98}. Radio
and optical data are nonsimultaneous and are taken from the archive
sources (see B{\l}a{\.z}ejowski \ea~1997 for details and references
therein).  Right panel: model fit to the outburst spectrum observed in
3C~279. Radio, optical, X--ray and $\gamma$--ray data are simultaneous
\citep{weh98} (Model A+B). \label{fig4}}

\figcaption[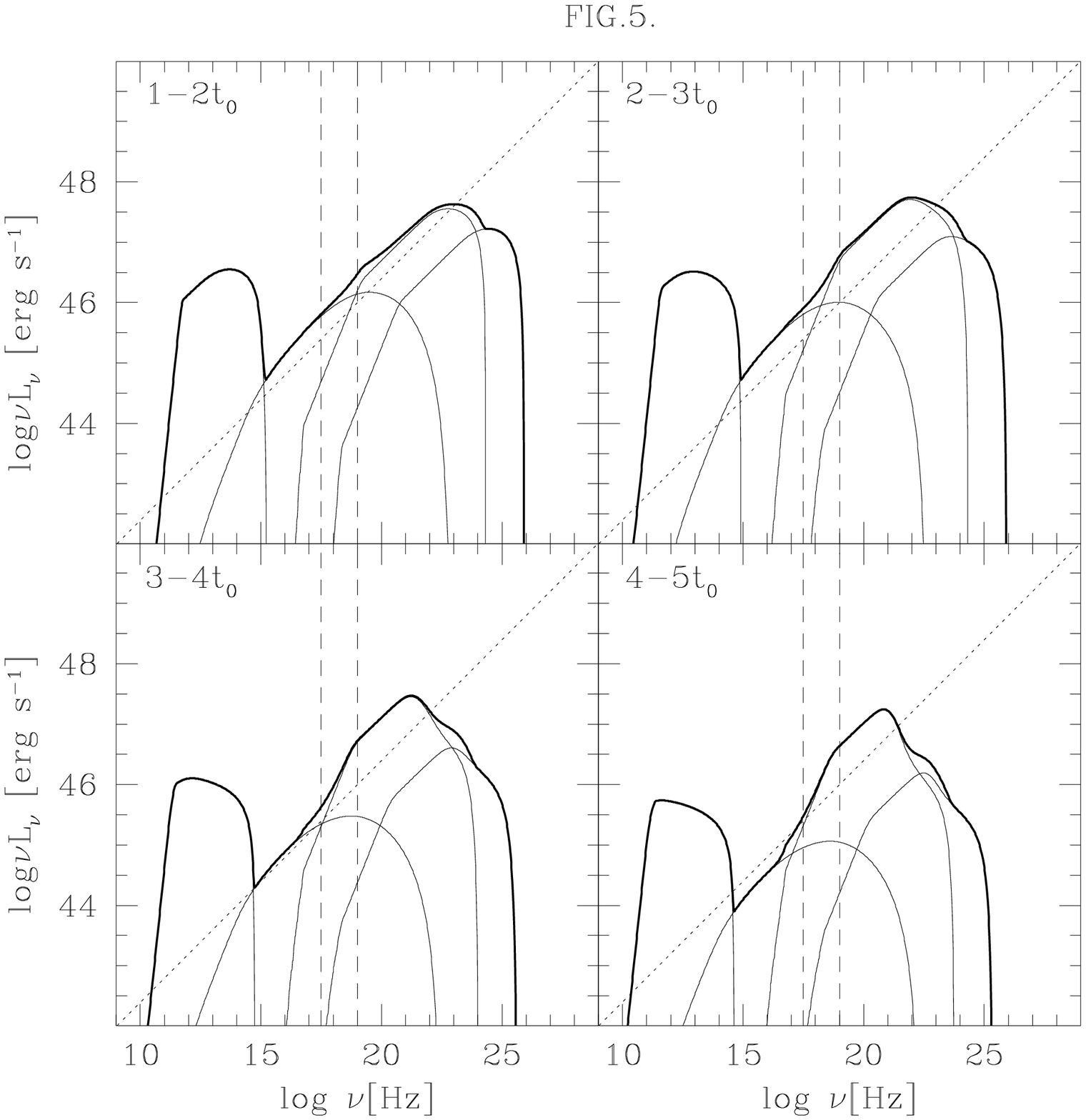]{Binned spectra in the time intervals $\Delta t = r_0 / 2
\Gamma^2 c$ (Model A+B). \label{fig5}}

\clearpage

\begin{deluxetable}{crrrrrrrrr}
\tabletypesize{\scriptsize}
\tablecaption{Parameters used in models \label{tab1}}
\tablewidth{0pt}
\tablehead{
\colhead{MODEL} & \colhead{${\Gamma}$} & \colhead{$r_0 [cm]$} &
\colhead{$B_0 [G]$} & \colhead{$K [{1 \over s}]$} &
\colhead{${\gamma}_b$} & \colhead{${\gamma}_{max}$} & \colhead{p} &
\colhead{$U_{BEL,0} [{erg \over {cm^3}}]$} & \colhead{$U_{IR} [{erg
\over {cm^3}}]$}
}
\startdata
A &$15$ &$7.0 \times 10^{17}$ &$0.43$ &$6.8 \times 10^{49}$ &$1-100$ &$6.0
\times 10^3$ &$2.2$ &$0.0$ &$10^{-4}$ \\
B &$15$ &$6.0 \times 10^{17}$ &$1.34$ &$6.9 \times 10^{49}$ &$1-100$ &$1.0
\times 10^4$ &$2.2$ &$2.3 \times 10^{-3}$ &$0.0$ \\
A+B &$15$ &$7.0 \times 10^{17}$ &$0.43$ &$8.9 \times 10^{49}$ &$1-100$ &$6.0
\times 10^3$ &$2.2$ &$ 10^{-4}$ &$10^{-4}$ \\
\enddata
\end{deluxetable}
\label{blaz_tab1}


\begin{deluxetable}{crrrrrrrrr}
\tabletypesize{\scriptsize}
\tablecaption{Parameters used in modeling 3C 279 and PKS 0528+134 \label{tab2}}
\tablewidth{0pt}
\tablehead{
\colhead{OBJECT} & \colhead{${\Gamma}$} & \colhead{$r_0 [cm]$} &
\colhead{$B_0 [G]$} & \colhead{$K [{1 \over s}]$} &
\colhead{${\gamma}_b$} & \colhead{${\gamma}_{max}$} & \colhead{p} &
\colhead{$U_{BEL,0} [{erg \over {cm^3}}]$} & \colhead{$U_{IR} [{erg
\over {cm^3}}]$}
}
\startdata
3C 279 &20 &${7.0 \times 10^{17}}$ &0.81 &${2.8 \times 10^{50}}$ &150 &$6.5
\times 10^3$ &2.4 &${0.6 \times 10^{-3}}$ &${0.2 \times 10^{-3}}$ \\
PKS 0528+134 &15 &${7.0 \times 10^{17}}$ &1.7 &${1.2 \times 10^{50}}$ &25 &$4.7
\times 10^3$ &2.2 &${2.1 \times 10^{-3}}$ &${4.7 \times 10^{-4}}$ \\
\enddata
\end{deluxetable}


\begin{deluxetable}{crrrrr}
\tabletypesize{\scriptsize}
\tablecaption{$(n_+/n_p)_{min}$ ratios. \label{tab4}}
\tablewidth{0pt}
\tablehead{
\colhead{${\gamma _b}$} & \colhead{1} & \colhead{5} & \colhead{20} &
\colhead{100}
}
\startdata
A &96 &40 &12 &2 \\
B &97 &41 &12 &2 \\
\enddata
\end{deluxetable}

\clearpage

\centerline{\psfig{file=blaz_fig1.ps,height=5.3 in,angle=0}}
\vfill\eject
\centerline{\psfig{file=blaz_fig2.ps,height=5.3 in,angle=0}}
\vfill\eject
\centerline{\psfig{file=blaz_fig3.ps,height=5.3 in,angle=0}}
\vfill\eject
\centerline{\psfig{file=blaz_fig4.ps,height=5.3 in,angle=0}}
\vfill\eject
\centerline{\psfig{file=blaz_fig5.ps,height=5.3 in,angle=0}}

\end{document}